\documentclass[onecolumn,usenatbib,amsmath,amssymb,amsbsy]{mn2e}

\usepackage{natbib}
\usepackage{graphicx}

\newcommand{\be}{\begin{equation}}
\newcommand{\ee}{\end{equation}}
\newcommand{\bear}{\begin{eqnarray}}
\newcommand{\eear}{\end{eqnarray}}
\newcommand{\dpe}{\delta p}

\begin{document} 

\title{Ekman layer damping of r-modes revisited}
  
\author[Glampedakis \& Andersson]{Kostas Glampedakis and Nils Andersson \\
School of Mathematics, University of
Southampton, Southampton SO17 1BJ, UK}

\maketitle

\begin{abstract}
We investigate the damping of neutron star r-modes due to the
presence of a viscous boundary (Ekman) layer at the interface between the crust and the core. Our study is motivated by 
the possibility that the gravitational-wave driven instability of the 
inertial r-modes may become active in rapidly spinning neutron stars, eg. in 
low-mass X-ray binaries, and the fact that a viscous Ekman layer at the 
core-crust interface  provides an efficient damping mechanism 
for these oscillations. 
We review various approaches to the problem and carry out an analytic 
calculation of the effects due to the Ekman layer for a rigid crust. 
Our analytic estimates support previous numerical results, 
and provide further insight into the intricacies of
the problem. We add to previous work by discussing 
the effect that compressibility and 
composition stratification have on the boundary layer damping. We show that, 
while stratification is unimportant for the r-mode problem, composition suppresses 
the damping rate by about a factor of two (depending on the detailed equation of state).
\end{abstract}


\section{Introduction}
\label{sec:intro}

During the past few years we have  witnessed a renewed interest 
in neutron star oscillations. One trigger for the recent activity
was the discovery that
inertial modes in rotating fluid stars are generically unstable
under the influence of gravitational radiation (a specific family of
 axial inertial modes, the so-called r-modes, being the most relevant)
\citep{na98,morsink}. Potential astrophysical implications of this 
instability have been extensively discussed in the literature, particularly in 
the context of gravitational wave observations 
(for a detailed review  see \citet{na_rev}). 
Another motivation for work in this area is provided by the possible  
observations of crustal oscillations following magnetar flares \citep{israel,watts}.
 
Soon after the discovery of the r-mode instability, 
it became clear that many additional pieces of
physics need to be considered if one wants 
to draw ''reliable'' astrophysical conclusions. Since much of the
required physics is poorly, or only partially,
known this effort is ongoing. At the present time the most important 
damping mechanisms that counteract the growth of an unstable mode
are thought to be i) viscous boundary layers at the core-crust 
interface \citep{BU} and their magnetic analogue \citep{mend1,kinney}, ii)
the bulk viscosity due to weak interactions in a hyperon core 
\citep{lind1,owen2}, and iii) superfluid mutual friction due to scattering 
of electrons off of  the vortices in the neutron superfluid 
in the core \citep{lind2,sidery} . 
At the same time the mode amplitude is expected
to saturate due to nonlinear mode-mode coupling \citep{arras,brink}.     
As the  damping of unstable r-modes could be 
an issue of astrophysical relevance these mechanisms have 
attracted some attention. 

At the time of writing, the available results indicate that for mature neutron stars the 
key damping agent may be 
the friction associated with the crust interface. 
The crust forms when the star cools
below $10^{10}$~K or so,  a few weeks to months after the star is born.
Shortly after this the heavy constituents in the core (neutrons, protons, 
hyperons) are expected to form superfluid/superconducting states. 
This leads to the suppression of, for example, the hyperon bulk viscosity \citep{haensel,reis,owen2}
and the emergence of mutual friction. The estimates for superfluid mutual friction \citep{lind2}
indicate that, even though this mechanism may suppress the instability in the f-modes completely \citep{lindlim}, 
it is not very efficient for r-modes \citep{lindmen}.
 
We therefore focus our attention on the role of the crust, which 
as a first approximation can be 
modelled as a rigid spherical ``container'' enclosing the neutron star fluid.
Such configurations have been studied in several classic fluid dynamics 
papers (see the standard textbook by \citet{Greenspan} for a detailed 
discussion of rotating fluids and a wealth of references). 
In the context of r-mode damping the effect of a rigid 
crust was first discussed 
some years ago by \citet{BU},  followed by the work of \citet{na1},  \citet{rieutord} and 
\citet{owen}. In this problem the core fluid oscillations are damped because of the 
formation of an Ekman layer at the base of the crust and the associated Ekman circulation. 
The resulting damping is rapid enough 
to  compete effectively with the gravitational-wave driving of the r-modes. 
Subsequent work by \citet{mend1} indicates that when a magnetic field 
is present in the Ekman layer the  damping timescale may become even shorter.
Thus the dissipation due to the presence of the crust-core interface
plays a key role for the r-mode instability. 

Our aim in this paper is to study the interaction between the oscillating fluid 
core and a rigid crust (using mainly analytical tools). 
A more refined model, which considers an elastic crust, is discussed in an accompanying paper (see also \citet{LU}).
Even though this problem has been discussed in some detail already, we have good motivation for 
revisiting it. It would be a mistake to 
think that the effect of the Ekman layer on an unstable r-mode is well understood. 
While we may know the answer for the simplest model problem of a viscous fluid in a (more or less) rigid
container, a number of issues remain to be investigated for realistic neutron stars. 
The presence of multiple particle species and the associated composition gradients
may affect the Ekman circulation \citep{abney}, the magnetic field boundary layer \citep{mend1} depends strongly
on the internal field structure (which is largely unknown) and whether the core forms a
superconductor or not, once the core is superfluid one would need to account for 
multifluid dynamics as well as possible vortex pinning, and so on. To prepare the ground for an assault on the 
latter two problems, we want to  obtain a better understanding of the nature of the 
viscous boundary layer and the methods used to study the induced damping of an oscillation mode. 

We will focus on the problem of a single r-mode in the core (a slight restriction, but
the generalisation to other cases is not very difficult once the method is developed). In order for
an analytic calculation to be tractable, a simple model for a neutron star is a
prerequisite. Hence, we first consider a uniform density, slowly rotating, Newtonian star
and reproduce familiar results for the Ekman problem. In doing this, we develop a
new scheme which should prove useful in more complicated problem settings.  
Our study also serves as an introduction to 
boundary layer techniques for the reader who is not already familiar with
the tricks of the trade. We then compare and contrast the different 
methods that have been used to investigate the Ekman layer problem. This discussion serves to clarify 
the current understanding of the induced r-mode damping rate. Finally, we extend our analysis
to compressible fluids with internal stratification.  The results of \citet{abney}
for the spin-up problem suggest that both these features will have a strong effect on the 
Ekman layer damping. However, we prove this expectation to be wrong. For a toroidal mode, like the r-mode, 
we find that
stratification is  irrelevant. The compressibility, on the other hand, does play a role
but we demonstrate that the effect is much weaker than indicated by \citet{abney}. 
Rather than leading to an exponential suppression, the
compressibility is estimated to reduce the r-mode damping by about a factor of two.


\section{Formulating the Ekman problem in a sphere}
\label{sec:Ekman}

As a first application of our boundary layer scheme
we consider the Ekman layer problem in a spherical container. This is an
 instructive exercise as the solution to this problem is well known \citep{Greenspan}. 
The so-called ``spin-up problem'' forms a part of classic fluid dynamics 
and was studied in a number of papers in the 1960s and 70s. We found the work by 
\citet{gh,green,clark,ache, bent} particularly useful.
The Ekman problem is also relevant for a 
discussion of unstable r-modes. In fact, it is reasonable to expect that the 
associated r-mode damping timescale is of the same order
of magnitude as the spin-up timescale in a rigid sphere since an inertial mode has frequency
proportional to the rotation rate of the star,  $\sigma \propto \Omega$. 

In building a simple model of a neutron star with a rigid crust, we 
assume a core consisting of an incompressible, uniform density, viscous fluid surrounded by
a rigid boundary. Furthermore, we work in a slow-rotation framework (which greatly
simplifies the mathematical analysis as the star is retains its spherical 
shape at ${\cal O}(\Omega)$). Finally, we  make the so-called Cowling 
approximation, i.e. neglect any perturbation in the gravitational potential. 
These simplifications are obviously not necessary in a numerical analysis of the problem, but 
(in our view) the benefits of having an analytic solution far outweighs 
the value of a numerical solution to the complete problem. One can argue in favour 
of some of our assumptions, since most astrophysical neutron stars are 
slowly rotating (in the sense that their spin is far below the break-up rate) 
and the inertial modes which we are interested in are such that the density 
(and hence gravitational potential) perturbations are higher order corrections.   
  
Working in the rotating frame, our dynamical equations are the (linearised) 
Euler and continuity equations, 
\bear
&& \partial_t {\bf v} + 2 {\bf\Omega} \times {\bf v} =  
- \nabla h + \nu \nabla^2 {\bf v}
\label{navier}
\\
&& \nabla \cdot{\bf v} = 0
\label{cont0}
\eear 
The enthalpy $h$, which is defined as
\be
\nabla h = { 1 \over \rho} \nabla p
\ee 
has been introduced for later convenience.
In addition to the equations of motion, we have two boundary conditions for the core flow: 
(i) regularity at $r=0$ and (ii) a ``no-slip'' condition ${\bf v}(R_c) = 0$ 
at the crust-core interface $r=R_c$. The role of viscosity is to ensure that 
the second condition is satisfied by modifying the inviscid solution in a thin 
boundary layer near the crust. This is known as the Ekman layer, and it has characteristic 
width $\delta_\mathrm{E} \approx \sqrt{ \nu /\Omega} \sim 1$~cm for typical neutron star parameters. 
Our aim in the 
following is to find a consistent solution to the problem when the solution 
in the bulk of the interior (the inviscid flow) corresponds to a {\it single} $l=m$ r-mode.

Schematically we assume that the solution takes the following form
(a standard uniform expansion in boundary layer theory \citep{orszag}):
\be
{\bf v} = {\bf v}^0 + \tilde{\bf v}^0  + \nu^{1/2} [ {\bf v}^1  + 
\tilde{\bf v}^1 ] + {\cal O}(\nu)
\label{uniform1}
\ee
and similarly for the perturbed enthalpy $\delta h$ and the mode frequency $\sigma$ (we assume
that all perturbations behave as $\exp(i\sigma t)$) 
\bear
\delta h &=& \delta h^0 +  \delta \tilde{h}^0 + \nu^{1/2} [ \delta h^1
+ \delta \tilde{h}^1 ] + {\cal O}(\nu)
\label{uniform2}
\\
\nonumber \\
\sigma &=& \sigma_0 + \nu^{1/2} i s  + {\cal O}(\nu)
\label{uniform3}  
\eear
The complex-valued frequency correction $s$ is of particular importance, since
its real part corresponds to the Ekman layer dissipation rate.
The idea behind the expansion (\ref{uniform1}) is that the inviscid solution 
${\bf v}^0$ is modified by  $\tilde{\bf v}^0$ in the boundary layer in such a way that 
the no-slip condition is satisfied. This then leads to a change in the
interior motion, represented by ${\bf v}^1$, which requires another
boundary layer correction $\tilde{\bf v}^1$ and so on. In principle, this leads to an infinite
hierarchy of coupled problems. In this
picture, the various boundary/crust-induced corrections to the basic inviscid flow
fall into two categories, namely, boundary-layer corrections (always 
denoted by tildes in this paper) which are characterised by rapid (radial) 
variation  [here $\partial_r \sim {\cal O}(\nu^{-1/2})$] and secondary 
``background'' corrections which vary smoothly throughout the star. Based on this
different behaviour, when (\ref{uniform1}) is inserted in (\ref{navier}) 
and (\ref{cont0}), each of these equations splits into two components;
one containing only boundary layer quantities and another comprising the background 
quantities. These two families of equations are not entirely independent, 
since the boundary conditions communicate information between 
boundary layer and background quantities.    


Assuming that the solution to the inviscid problem takes the form of 
an axial r-mode we get
\begin{eqnarray}
\partial_t v^0_\theta - 2 \Omega \cos \theta v^0_\varphi &=& - 
{1 \over r}\partial_\theta \delta h^0 \\
\partial_t v^0_\varphi + 2 \Omega \cos \theta v^0_\theta 
&=& - {1 \over r \sin \theta }\partial_\varphi \delta h^0 \\
 2 \Omega \sin \theta v^0_\varphi &=& \partial_r \delta h^0
\label{modedec}\end{eqnarray}
We also know that the $l=m$ r-mode solution corresponds to a frequency
\be
\sigma_0 = { 2 \Omega \over m+1 }
\ee
and can be expressed in terms of a stream function $U^0$ as
\be
{\bf v}^0 = - {  \hat{\bf e}_\theta \over r \sin \theta} i \partial_\varphi U^0
+ { \hat{\bf e}_\varphi \over r} i \partial_\theta U^0 
\ee
where the phase is chose for later convenience, and 
\be
U^0 = U_m^0 Y_{l=m}^m(\theta,\varphi) e^{i\sigma t} \ , \quad \mbox{ with } \quad  U_m^0 = \left( {r \over R_c} \right)^{m+1} 
\label{rfun}\ee 
(normalised to unity at the base of the crust).

The corresponding boundary layer correction is a solution to 
\begin{eqnarray}
\partial_t \tilde{v}^0_\theta - 2 \Omega \cos \theta \tilde{v}^0_\varphi 
&=& \nu \partial_r^2 \tilde{v}^0_\theta \label{vbl1} \\
\partial_t \tilde{v}^0_\varphi + 2 \Omega \cos \theta 
\tilde{v}^0_\theta &=& \nu \partial_r^2 \tilde{v}^0_\varphi \label{vbl2}\\
 2 \Omega \sin \theta \tilde{v}^0_\varphi &=&
\partial_r \delta \tilde{h}^1 \\
\partial_r \delta \tilde{h}^0 &=& 0 
\end{eqnarray}
We also need to satisfy the continuity equation which leads to 
$\tilde{v}^0_r=0$, and
\be
 \nu^{1/2} r \partial_r \tilde{v}^1_r + { 1 \over \sin \theta} \partial_\theta
(\sin \theta \tilde{v}^0_\theta) + 
{ 1 \over \sin \theta}  \partial_\varphi \tilde{v}^0_\varphi = 0 \label{vbl3}
\ee
Note that we have only kept the highest radial derivative of the various
variables in the boundary layer, and that we have also neglected angular
derivatives compared to radial ones. 
Finally, the solution to the $\tilde{\bf v}^0$ problem must be such that
the no-slip boundary condition is satisfied, i.e. we must have
\be
{\bf v}^0 + \tilde{\bf v}^0 = 0 \ , 
\qquad \hbox{ at } r =R_c \ .
\ee 

At the next level, we determine the change in the 
interior motion induced by the presence of the Ekman layer. 
This means that we must solve,
\begin{eqnarray}
\partial_t v^1_\theta - 2 \Omega \cos \theta v^1_\varphi 
&=& - { 1 \over r} \partial_\theta \delta h^1 + s v^0_\theta 
\label{int1}\\
\partial_t v^1_\varphi + 2 \Omega \cos \theta 
v^1_\theta + 2 \Omega \sin\theta v^1_r &=&- { 1 \over r \sin \theta} 
\partial_\varphi \delta h^1 + s v^0_\varphi  \label{int2}\\
\partial_t v^1_r - 2 \Omega \sin \theta v^1_\varphi &=& - 
\partial_r \delta h^1 
\label{induced}
\end{eqnarray}
together with the continuity equation
\be
\nabla \cdot {\bf v}^1 = 0 \label{int3}
\ee
and a boundary condition
\be
v^1_r + \tilde{v}^1_r = 0
\ee
A solution to this problem should provide the frequency correction $s$, which encodes the rate at which the
presence of the Ekman layer alters the core flow. That is, the induced damping of the r-mode
which we are interested in.


\section{Solution in terms of a spherical harmonics expansion}
\label{sec:spherical} 

So far, our calculation follows closely the standard treatment of
the spin-up problem \citep{gh,Greenspan}. At this point we deviate, by expanding the velocity 
components in spherical harmonics. We have several reasons for doing so. Firstly, the 
standard approach (which will be reviewed in Section~\ref{sec:standard}) leads to a boundary layer flow 
that is explicitly singular in particular angular directions. Although one can demonstrate that 
this is not a great problem, and that the predicted Ekman layer dissipation can be 
trusted, it is not an attractive feature of the solution. 
We want to use an approach that avoids such singular behaviour. 
Secondly, the standard approach is somewhat limited in that it may not easily allow 
us to consider a differentially rotating background flow
(or an interior magnetic field with complex multipolar structure). 
Our approach will allow such generalisations, should they be required.


\subsection{The boundary layer corrections}

We first concentrate on the correction to the fluid velocity in the 
Ekman layer. We need to solve (\ref{vbl1})-(\ref{vbl2}) and then 
determine $\tilde{v}^1_r$ from (\ref{vbl3}). If we are mainly interested 
in determining the damping timescale due to the presence 
of the Ekman layer, the induced radial flow will play the 
main role. 

Assuming that $\tilde{\bf v}^0$ can be written
\be
\tilde{\bf v}^0 = { \tilde{W} \over r} \hat{\bf e}_r + { 1 \over r}
\left( \partial_\theta \tilde{V} - { i \partial_\varphi \tilde{U} \over
\sin \theta} \right)\hat{\bf e}_\theta +   { 1 \over r}
\left( {\partial_\varphi \tilde{V} \over \sin \theta} + i \partial_\theta 
\tilde{U} \right)\hat{\bf e}_\varphi
\label{stream}
\ee
i.e. as a general decomposition into polar ($\tilde{W}, \tilde{V}$) and axial ($\tilde{U}$) perturbations, 
and that our solution depends on $\varphi$ as $e^{im\varphi}$, we get from 
(\ref{vbl1}) 
\be
m \cos \theta \tilde{V} + \sin \theta \cos \theta \partial_\theta \tilde{U}
= { 1 \over 2\Omega} (\sigma_0 + i\nu \partial^2_r) (\sin \theta \partial_\theta 
\tilde{V} 
+ m \tilde{U})
\label{vbl4}\ee
Similarly, equation (\ref{vbl2}) leads to 
\be
 \sin \theta \cos \theta \partial_\theta \tilde{V} +
m \cos \theta \tilde{U} 
= { 1 \over 2\Omega} (\sigma_0 + i\nu \partial^2_r) (m \tilde{V} + 
\sin \theta \partial_\theta \tilde{U} )
\label{vbl5}\ee
 
We can  combine these two equations to get
\begin{eqnarray}
- { 1 \over 2\Omega} ( \sigma_0 +i\nu \partial^2_r) \nabla^2_\theta 
\tilde{U} &=& m \tilde{U} - \cos \theta  \nabla^2_\theta \tilde{V} 
+ \sin \theta \partial_\theta \tilde{V} \\
{ 1 \over 2\Omega} ( \sigma_0 +i\nu \partial^2_r) \nabla^2_\theta 
\tilde{V} &=& - m \tilde{V} + \cos \theta  \nabla^2_\theta \tilde{U} 
- \sin \theta \partial_\theta \tilde{U}
\end{eqnarray}
where 
\be
\nabla^2_\theta = { 1 \over \sin \theta} \partial_\theta ( \sin \theta 
\partial_\theta ) + { \partial_\varphi^2 \over \sin^2 \theta} 
\ee
is the angular part of the Laplacian on the unit sphere. 

If we now expand the various functions in spherical harmonics, i.e. $\tilde{U} = \sum_l \tilde{U}_l Y_l^m$
etc, and then use their orthogonality together with the
standard recurrence relations,
\begin{eqnarray}
\cos \theta Y_l^m &=& Q_{l+1} Y_{l+1}^m + Q_l Y_{l-1}^m \\
\sin \theta \partial_\theta Y_l^m &=& l Q_{l+1}Y_{l+1}^m - (l+1) Q_l Y_{l-1}^m
\label{Ylm_id}
\end{eqnarray}
where 
\be
Q_l = \sqrt{\frac{(l+m)(l-m)}{(2l-1)(2l+1)}}
\ee
we arrive at the two coupled equations
\begin{eqnarray}
{ n(n+1) \over 2\Omega} ( \sigma_0 + i\nu \partial^2_r ) \tilde{U}_n 
- m \tilde{U}_n - (n-1)(n+1) Q_n \tilde{V}_{n-1} -
n(n+2)Q_{n+1} \tilde{V}_{n+1} &=& 0 \label{Ueq}\\
{ (n+1)(n+2) \over 2\Omega} ( \sigma_0 + i\nu \partial^2_r ) \tilde{V}_{n+1} 
- m \tilde{V}_{n+1} 
- n(n+2) Q_{n+1} \tilde{U}_n -
 (n+1)(n+3)Q_{n+2} \tilde{U}_{n+2} &=& 0
\label{Veq}\end{eqnarray}

Since we know that we must match our solution to the inviscid r-mode solution 
at $r=R_c$, we now {\it assume} that all toroidal components apart
from $\tilde{U}_m$ vanish. Strictly speaking, this assumption is
inconsistent. This can be easily seen by setting $n=m \to m+2$ 
in the first of equations above. In principle, one would expect an infinite sum of 
contributions. This problem is similar to that for an inertial mode in a 
compressible star \citep{lockitch}. 
From a practical point of view, however, one must truncate the sum at some level in 
order to close the system of equations. Just like in the inertial-mode problem,
one would expect the solution to converge as further multipole contributions are 
accounted for. This then provides a simple way to estimate the error induced by the 
truncation. Adopting this strategy, we will later compare our results to ones obtained 
by including  more axial terms ($\tilde{U}_{m+2}$, $\tilde{U}_{m+4}$ and so on) in the boundary layer flow.

After setting $n=m$ we get (since $Q_m=0$)
\begin{eqnarray}
{ m(m+1) \over 2\Omega} ( \sigma_0 + i\nu \partial^2_r ) \tilde{U}_m 
- m \tilde{U}_m -
m(m+2)Q_{m+1} \tilde{V}_{m+1} &=& 0 \\
{ (m+1)(m+2) \over 2\Omega} ( \sigma_0 + i\nu \partial^2_r ) \tilde{V}_{m+1} 
- m \tilde{V}_{m+1} 
- m(m+2) Q_{m+1} \tilde{U}_m  &=& 0
\end{eqnarray}

Using the standard ansatz $[\tilde{U}_m, \tilde{V}_{m+1}] \propto e^{\lambda r}$
for a differential equation with constant coefficients,
we can show that a solution will exist if
\be
\nu^2 \lambda^4 - { 4 i  \Omega \over (m+1)(m+2)}\nu \lambda^2
+ { 4m(m+2)\Omega^2 Q_{m+1}^2 \over (m+1)^2 }= 0
\ee
i.e. for the two roots
\be
\nu \lambda^2 = { 2i\Omega \over (m+1)(m+2)} (1\pm \alpha)
\label{lasq}\ee
where we have defined
\be
\alpha = \sqrt{m(m+2)^3Q_{m+1}^2+1 }
\ee

The solutions correspond to eigenvectors 
\be
\tilde{V}_{m+1} = - { 1\pm\alpha \over (m+2)^2 Q_{m+1}} \tilde{U}_m
\ee

In order to satisfy the prescribed boundary conditions at the 
crust-core interface, we must have a vanishing velocity at $r=R_c$. 
Writing the solution as
\be
\tilde{U}_m = A e^{\lambda_1(r-R_c)} + B e^{\lambda_2(r-R_c)}
\ee
where 
\be
\lambda_{1,2} = (1 \pm i) \sqrt{ (1+\alpha)\Omega \over \nu (m+1)(m+2) } 
\ee
we must have
\be
\tilde{U}_m (R_c) = U_m^0(R_c) \longrightarrow A+B=1
\ee
Furthermore, we also require that 
\be
\tilde{V}_{m+1} (R_c) = - { 1 \over (m+2)^2Q_{m+1}}
\left[ (1+\alpha)A + (1-\alpha)B \right] = 0
\ee
which leads to
\be
A = -  { 1-\alpha \over 2\alpha} \qquad \hbox{ and } 
\qquad B =  { 1+\alpha \over 2\alpha} 
\ee
  
Finally, we can use this solution together with (\ref{vbl3})
to determine $\tilde{v}^1_r$. Assuming that $\tilde{\bf v}^1$ 
also
takes the form (\ref{stream}), we readily find that
\be
\nu^{1/2} \partial_r (r \tilde{W}_{m+1}^1) = (m+1)(m+2) \tilde{V}_{m+1}
\label{rad1}
\ee
which can be  integrated to give $\tilde{W}_{m+1}^1$. Hence, we have
obtained all leading-order boundary layer variables.


\subsection{The induced interior flow}
\label{sec:induced}

Having determined the appropriate solution in the Ekman layer, we now
want to calculate the induced flow in the interior and deduce the
damping rate of the r-mode. By combining (\ref{int1}) and (\ref{int2}), 
and again using a solution of form (\ref{stream}) [albeit 
now without the tildes] for ${\bf v}^1$, we can show that we must have
\be
( \sigma_0 \nabla^2_\theta U + 2m\Omega U ) - 2\Omega \cos \theta 
\nabla^2_\theta V + 2\Omega \sin \theta \partial_\theta V -2\Omega [ \sin \theta \partial_\theta W + 2\cos\theta W] = 
- is \nabla^2_\theta U^0
\ee
Expanding once again in spherical harmonics and using the fact that the 
right-hand side vanishes unless $l=m$, we immediately 
arrive at 
\be
2 (m+2)\Omega Q_{m+1} V_{m+1} + 2\Omega Q_{m+1} W_{m+1} = 
i s (m+1) U^0_m
\ee
and a solution
\be
V_{m+1} = { s(m+1) \over 2\Omega (m+2) Q_{m+1}} 
U^0_m  -\frac{1}{m+2} W_{m+1}
\label{indV}\ee
Given this result, we can use the continuity equation to show that
\be
\partial_r(r W_{m+1}) = (m+1)(m+2) V_{m+1} = { is(m+1)^2 \over 2\Omega 
Q_{m+1}} U_m^0 -(m+1) W_{m+1}
\label{rad2}
\ee
(Note that the integration constant must vanish since we would otherwise have a
solution that diverges at the origin.)   

The final step in our analysis corresponds to combining (\ref{rad1}) 
and (\ref{rad2}) in such a way that 
\be
\tilde{W}_{m+1}(R_c)+ W_{m+1}(R_c) = 0 
\ee

By integrating the two solutions we find that 
\be
\nu^{1/2} R_c\tilde{W}_{m+1} = - { m+1 \over m+2} { 1 \over 2  Q_{m+1} }
{ \alpha^2 - 1 \over \alpha} \left[ { 1 \over \lambda_1} - 
{ 1 \over \lambda_2} \right]
\ee
and
\be
 W_{m+1}(R_c) = {(m+1)^2 i s Q_{m+1} \over 2 \Omega }
\label{wrad}\ee
After combining these and reshuffling, we have 
\be
s = {  \Omega \over \nu^{1/2} R_c} {(2m+3) \over (m+2)(m+1)} 
{i(\alpha^2  - 1) \over \alpha}
\left[ { 1 \over \lambda_1} - 
{ 1 \over \lambda_2} \right]
\label{damp}
\ee
From this last equation we can extract the damping rate,
\be
\hbox{Re s} = \left( {\Omega \over R_c^2} \right)^{1/2}  
{2m+3 \over \sqrt{ 4( m+1)(m+2)}} { \sqrt{\alpha+1} + 
\sqrt{\alpha-1} \over \alpha} 
\sqrt{ \alpha^2-1}
\ee
We also find that the frequency correction due to the presence of the
Ekman layer is
\be
\hbox{Im s} = \left( {\Omega \over R_c^2} \right)^{1/2}  
{2m+3 \over \sqrt{ 4(m+2)( m+1)}} 
{ \sqrt{\alpha-1} - \sqrt{\alpha+1} \over \alpha} 
\sqrt{ \alpha^2-1}
\ee
These expressions constitute the main new result of the first part of this paper. 
They have been derived by combining boundary-layer theory techniques with a 
decomposition in spherical harmonics. Putting numbers in, we find for $m=1$ that 
$s\approx 2.92 -0.602i$. This should be compared to the results of 
\citet{Greenspan} (see also \citet{rieutord}), who 
finds $s \approx 2.62 -0.259i$. The quantity which we are mainly interested in, the 
real part of $s$ which gives the damping rate of the core flow, 
differs from Greenspan's result by about 11\%. This is not too bad given that we truncated the 
boundary layer solution at the level of a single multipole contribution. 
If we include further terms, we need to solve a system of equations 
larger than the ``$2\times2$'' problem given by (\ref{Ueq})-(\ref{Veq}). 
Including one more axial term ($\tilde{U}_{m+2}$) in the boundary layer
we end up with a ``$3\times3$'' problem and so on. As we include more multipoles in 
the description our damping rate approaches Greenspan's result, cf. the results listed
in Table~\ref{table1}. We see that if we consider the ``$4\times4$'' 
problem we get results that agree with the standard ones, see the following section, 
to within 5\% or so. This is certainly much more accurate than our knowledge of the various
neutron star parameters. It should be noted that, by solving the problem in a
``non-standard'' way we have learned 
an interesting lesson: In order to represent the boundary layer flow 
accurately, we need to include a number of multipoles even though we
have a very simple core flow. 

\begin{table}
\begin{center}
\begin{tabular}{|c|c|c|c|}
\hline
 & truncation  & This work (Section~\ref{sec:spherical}) & Greenspan (Section~\ref{sec:standard}) \\
\hline 
$m=2$ & $2\times2$ & $4.097-0.473i$ & $3.518-0.176i$                   \\
      & $3\times3$ & $3.451+0.376i$ & \\
      & $4\times4$ & $3.714-0.292i$ & \\
\hline
$m=3$ & $2\times2$ & $5.067-0.391i$  & $4.288-0.136i$ \\
      & $3\times3$ & $3.922+0.592i$  & \\
      & $4\times4$ & $4.405+0.209i$  &  \\ 
\hline
\end{tabular}
\end{center}
\caption{Ekman dissipation rate $s$ computed in the present work,
compared with the standard approach in the literature \citep{Greenspan}.}
\label{table1}
\end{table}


\section{The standard solution to the Ekman problem}
\label{sec:standard}

For completeness, and comparison with our calculation, we will now discuss the ``traditional'' 
solution of the Ekman problem. This solution is described in detail in 
the paper by \citet{gh} as well as the book 
by \citet{Greenspan} and is, essentially, a boundary layer theory calculation 
without expanding in spherical harmonics. The formulation of the problem is identical to 
that described above. 

By combining (\ref{vbl1}) and (\ref{vbl2}) we can easily show that
we must have
\be
\left[ \nu^2 \partial_r^4 - 2i\sigma_0 \nu \partial_r^2 - \sigma_0^2 
+ 4\Omega^2 \cos^2 \theta \right]\tilde{v}_\theta = 0
\label{master1}
\ee
Despite really being a \underline{partial} differential equation, this is a now viewed as a ``constant coefficient''
\underline{ordinary} differential equation for $\tilde{v}_\theta$
as a function of $r$. The ansatz 
$\tilde{v}_\theta = e^{\lambda (r-R_c)}$ then
leads to the indicial equation
\be
\nu^2 \lambda^4 - 2i \sigma_0 \nu \lambda^2 + 4\Omega^2 \cos^2 \theta - \sigma_0^2 = 0
\ee
which has solutions
\be
\lambda = \pm \nu^{-1/2} \left[ i (\sigma_0 \pm 2 \Omega \cos \theta)\right]^{1/2}
\ee
In order for the boundary layer corrections to the velocity field to 
die away from the surface we must have $\mbox{Re } \lambda > 0$. 
One can show that the two solutions that satisfy this criterion 
correspond to 
\be
\bar{\lambda}_{1,2} = \left| { 1 \over m+1} \pm \cos \theta \right|^{1/2}
\left[ 1 + {1/(m+1) \pm \cos \theta \over |1/(m+1) \pm \cos \theta|} i \right] 
\ee
where we have i) used the $l=m$ r-mode frequency, and
ii) introduced $\lambda = \sqrt{\Omega/\nu}\, \bar{\lambda}$. 

Given a solution for $\tilde{v}_\theta$ we find from (\ref{vbl2}) that
\be
\pm i \tilde{v}_{\varphi} = \tilde{v}_\theta
\ee
where the upper (lower) sign corresponds to $\lambda_1$ ($\lambda_2$). 
The leading order correction to the velocity field due to the presence of the 
Ekman layer can now be written
\begin{eqnarray}
\tilde{v}_\theta &=& A e^{\lambda_1 (r-R_c)} + B e^{\lambda_2 (r-R_c)} \\ 
\tilde{v}_\varphi &=& -iA e^{\lambda_1 (r-R_c)} +i B e^{\lambda_2 (r-R_c)} 
\end{eqnarray}
The coefficients $A,B$ here should not be confused with those appearing
in Section~\ref{sec:spherical} (although in both cases they have similar roles). 
This solution needs to be matched to the inviscid solution in such a 
way that the no-slip condition at $r=R_c$ is satisfied. 
Matching to the $l=m$ inviscid r-mode leads to
\begin{eqnarray}
A &=& { m \over 2R_c \sin \theta} Y_m^m - {i  \over 2R_c} \partial_\theta Y_m^m \\
B &=& { m \over 2R_c \sin \theta} Y_m^m + {i  \over 2R_c} \partial_\theta Y_m^m 
\end{eqnarray}
This completes the solution in the boundary layer.
We want to infer the induced radial flow in order to estimate the 
Ekman layer damping of the core r-mode oscillation. The required 
relation is  best expressed in terms of a ``stretched'' boundary 
layer variable
\be
\zeta = \sqrt{ \Omega \over \nu} (R_c-r)
\ee 
Then the ${\cal O}(1)$ continuity equation leads to 
\be
\partial_\zeta \tilde{v}_r^1 = {1\over \sqrt{\Omega R_c^2} } 
\left[ { 1 \over \sin \theta} \partial_\theta (\sin \theta \tilde{v}_\theta)
+ {1 \over \sin \theta} \partial_\varphi \tilde{v}_\varphi \right] 
\equiv { {\cal I} \over  \sqrt{ \Omega R_c^2}}
\ee
or
\be
 \tilde{v}_r^1 = {1 \over\sqrt{ \Omega R_c^2}} \int_0^\zeta 
\left[ { 1 \over \sin \theta} \partial_\theta (\sin \theta \tilde{v}_\theta)
+ {1 \over \sin \theta} \partial_\varphi \tilde{v}_\varphi \right] d \zeta  =  
{ 1 \over  \sqrt{ \Omega R_c^2}} \int_0^\zeta {\cal I} d \zeta
\ee
This is to be matched to the $O(\nu^{1/2})$ core solution in such a way
that 
\be
\lim_{r\to R_c} v_r^1 = \lim_{\zeta \to \infty} \tilde{v}_r^1
\label{matchvr}
\ee

Working things out for our  boundary layer solution we find that
\be
\lim_{\zeta \to \infty} \tilde{v}_r^1 
= {1 \over \sqrt{ \Omega R_c^2}} \int_0^\infty {\cal I} d\zeta =  
{1 \over \sqrt{ \Omega R_c^2} }
\left\{  { i m(m+1) \over 2R_c} \left[ { 1 \over \bar{\lambda}_1} - 
{ 1 \over \bar{\lambda}_2}  \right] Y_m^m
- A { \partial_\theta \bar{\lambda}_1 \over \bar{\lambda}_1^2}
- B { \partial_\theta \bar{\lambda}_2 \over \bar{\lambda}_2^2}
\right\}
\ee

Using the induced core solution from before, i.e. (\ref{wrad}), we have
\be
\lim_{r\to R_c} v_r^1 = { (m+1)^2 s \over 2 (2m+3) R_c \Omega Q_{m+1}} Y_{m+1}^m
\ee
and we can determine the Ekman layer dissipation rate for the r-mode from
\be
{ (m+1)^2 s \over 2 (2m+3)  Q_{m+1}} Y_{m+1}^m = 
\sqrt{  \Omega \over R_c^2} \left\{  
 { i m(m+1) \over 2} \left[ { 1 \over \bar{\lambda}_1} - 
{ 1 \over \bar{\lambda}_2}  \right] Y_m^m
- AR_c { \partial_\theta \bar{\lambda}_1 \over \bar{\lambda}_1^2}
- BR_c { \partial_\theta \bar{\lambda}_2 \over \bar{\lambda}_2^2}
\right\}
\ee
The final step is to use orthogonality of the spherical harmonics, 
i.e. evaluate $s$ from
\be
s = { 2 (2m+3) Q_{m+1} \over (m+1)^2 } \sqrt{\frac{\Omega}{R_c^2}} {\cal J}
\label{firsts}\ee
where 
\be
{\cal J} = 2\pi \int_0^\pi \sin \theta  \bar{Y}_{m+1}^m \left\{ 
{ i m(m+1) \over 2} \left[ { 1 \over \bar{\lambda}_1} - 
{ 1 \over \bar{\lambda}_2}  \right] Y_m^m
- AR_c { \partial_\theta \bar{\lambda}_1 \over \bar{\lambda}_1^2}
- BR_c { \partial_\theta \bar{\lambda}_2 \over \bar{\lambda}_2^2}\right\} 
d\theta 
\ee
The required integral can be much simplified if we use the symmetry 
of the integrand, see the discussion in the Appendix or alternatively \citet{liao}. Thus we get
\be
{\cal J} =  { 2\pi i} \int_0^\pi { \sin \theta  \over \bar{\lambda}_1}
\left\{  { m^2 \over \sin^2 \theta} 
 \bar{Y}_{m+1}^m Y_m^m - { m \over \sin \theta} \partial_\theta (  
\bar{Y}_{m+1}^m Y_m^m ) + (\partial_\theta  \bar{Y}_{m+1}^m)(\partial_\theta 
Y_m^m)
\right\} d\theta
\label{Jint}\ee
The obtained values for the damping rate $s$ are given in Table~\ref{table1}.  

\begin{figure}
\centerline{\includegraphics[height=5cm,clip] {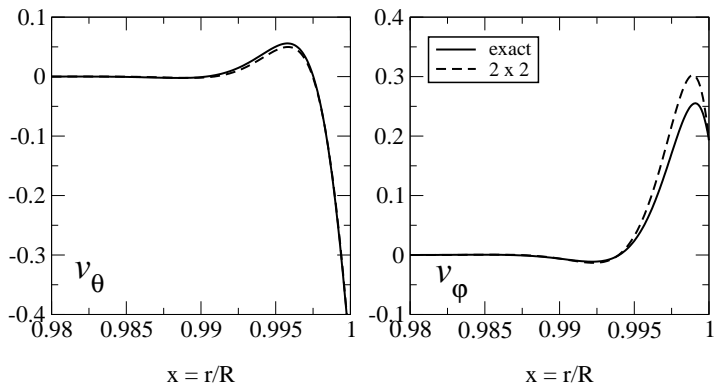}}
\caption{We compare the solution of the $2\times2$ problem (dashed) to the standard ``exact'' solution
(solid line). The data  corresponds to parameters $m=2$, $E=\nu/2\Omega R^2= 10^{-6}$,  $\varphi=\pi/3$
and a typical angle $\theta=\pi/4$. The two solutions are clearly in good agreement.}
\label{fig2}
\end{figure}

\begin{figure}
\centerline{\includegraphics[height=7cm,clip] {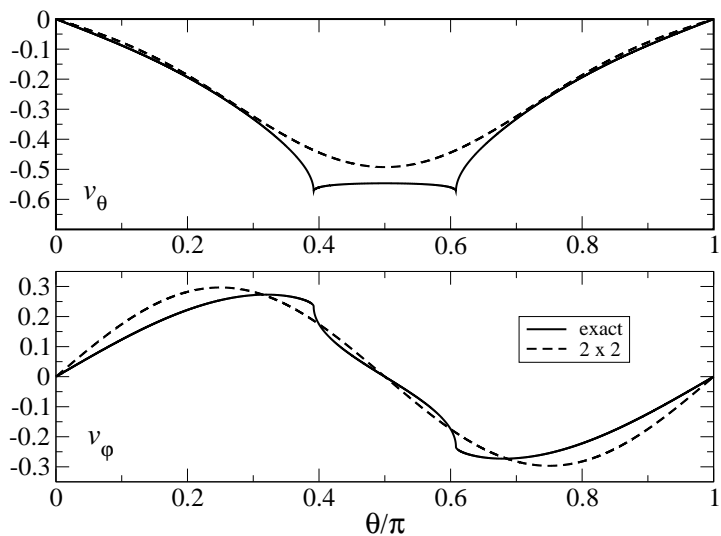}}
\caption{We compare the solution of the $2\times2$ problem (dashed) to the standard ``exact'' solution
(solid line). The data  corresponds to parameters $m=2$, $E=\nu/2\Omega R^2=10^{-6}$, $\varphi=\pi/3$
and a distance away from the boundary given by $x=r/R=1-0.8\sqrt{E}$. The particular values of $\theta$ for which the 
standard solution is singular are apparent in the figure. However, away from these angles we find the 
two solutions to agree well.}
\label{fig3}
\end{figure}

The solution to the Ekman problem  described in this Section is straightforward
but somewhat limited, in the sense that it relies on the simplicity of equation
(\ref{master1}). In a more general context, this approach may no longer be possible.
Moreover, as we already pointed out, this  solution exhibits a singular 
behaviour for certain angles (when $\lambda=0$). 
From an intuitive point of view one would clearly not expect a physical quantity, in this case the velocity 
field of the fluid, to exhibit singularities in certain angular directions.
Hence, it is useful to discuss these singularities 
in more detail.
To unveil the origin of the singular solution we need to return to 
 Eq. (\ref{navier}). Key to the boundary layer approach was the assumption that 
the viscous term on the right-hand side is relevant only in a narrow region in the vicinity of (in our case) the neutron star crust.  
The idea  is that the inviscid solution 
${\bf v}^0$ is modified by a correction
 $\tilde{\bf v}^0$ in the boundary layer in such a way that 
a no-slip condition can be imposed. The boundary-layer corrections 
are characterised by rapid (radial) 
variation  [$\partial_r \sim {\cal O}(\nu^{-1/2})$]. Thus, neglecting the 
angular derivatives in the Laplacian in the right-hand side of the Navier-Stokes 
equation we can derive Eqs (\ref{vbl1})-(\ref{vbl2}). If we then conveniently
``forget'' that we are actually solving partial differential equations and treat $\theta$ as a ``constant'' then we  
arrive at  (\ref{master1}).  At the end of the day, the solutions 
we  obtain are manifestly singular for particular values of $\theta$. 
The problem stems from the  assumption 
that
the radial derivatives dominate the angular derivatives in the boundary layer.
Without this assumption we would have to 
solve the original partial differential equation, the standard approach to which is (of course) separation of variables via an expansion in spherical harmonics (as
in our method in Section~\ref{sec:spherical}).  However, given the singular solution derived above it is very easy to show that 
\begin{equation}
{\partial \tilde{v}_\theta \over \partial r} \ll {\partial \tilde{v}_\theta \over \partial \theta}
\end{equation}
near the singular angular directions. Hence, this solution 
violates the  assumptions made in the derivation which
 renders the calculation inconsistent. This argument shows that   
this solution is unphysical in the vicinity 
of these angular directions. Similar conclusions have been drawn by \citet{rieval}
 and \citet{holker}. 
They consider the complete problem numerically and show that shear layers form in these specific 
angular directions. They also prove that the presence of these shear layers have a small 
effect on the damping rate. This is the key reason why both our method from Section~\ref{sec:spherical} and the ``exact'' singular 
solution derived above, where the detailed shear layers are neglected, lead to consistent and 
reliable estimates for the damping due to the Ekman layer. A comparison of the two velocity fields, cf.
Figure~\ref{fig2} and \ref{fig3}, shows that the agreement is quite good away from the singular directions. 
Hence, it is not very surprising that the two calculations lead to 
similar estimates for the 
r-mode damping. It is worth noting that \citet{holker} argues that the relatively slow convergence of the correction to the oscillation frequency, 
cf. Table~\ref{table1}, is due to the presence of the shear layers.  

\section{Another solution: surface integral approach}
\label{sec:surf}

There is yet another alternative for solving the Ekman problem, based
on direct integration of the Euler equation. This approach is discussed
in Greenspan's book \citep{Greenspan}. 
With slight modifications, this method was used by \citet{rieutord}
and \citet{liao} in their studies of inertial modes.

Returning to equation (\ref{induced}), which determines the induced
core flow, and reinstating the pressure and density instead of the enthalpy, we have
\be
i \sigma_0 {\bf v}^1 + 2\Omega \times {\bf v}^1 + { 1\over \rho} {\bf \nabla}
\delta p^1 = s {\bf v}^0
\ee
Multiplying  by the complex conjugate $\bar{\bf{ v}}^0$ (the conjugate is denoted by a bar),  and using the 
(conjugate of the) inviscid equation
\be
i \sigma_0 {\bf v}^0 + 2 \Omega \times {\bf v}^0 = { 1\over \rho} \nabla \delta p^0
\ee
as well as the continuity equation $\nabla \cdot {\bf v}^0=0$,
we obtain,
\be
\nabla \cdot \left[ \delta p^1 \bar{{\bf v}}^0 + \delta p^0 \bar{\bf v}^1 \right] =
\rho s |{\bf v}^0|^2
\ee
Integration of this relation over the volume ($V$) of the star and use of 
the divergence theorem gives,
\be
\int_S \hat{\bf r} \cdot  \left[ \delta p^1 \bar{{\bf v}}^0 + \bar{(\delta p^0)} 
{\bf v}^1 \right] dS = s \int_V \rho  |{\bf v}^0|^2 dV
\ee
where $S$ represents the surface at the core-crust interface.
We know that the inviscid r-mode problem is such that 
\be
{\bf \hat{r}} \cdot \bar{{\bf v}}^0 = 0
\ee
which means that we arrive at the final expression
\be
s = { \int_S \bar{(\delta p^0)} ( \hat{\bf r} \cdot {\bf v}^1) dS \over  
\int_V \rho  |{\bf v}^0|^2 dV}
\label{sint} 
\ee

Now we can combine (\ref{sint}) with the matching condition for the induced 
flow (\ref{matchvr}). This provides a connection to the calculation in the previous 
section. We can easily determine $\delta p^0$ for the inviscid flow. 
From the r-mode solution to the velocity field we find that
\be
\delta p^0 = -i {2m \over m+1} \rho \Omega \left( {r \over R_c} \right)^{m+1} Q_{m+1} Y_{m+1}
\ee
To determine the damping rate we also need the volume 
integral (essentially the mode energy)
\be
\int_V \rho |{\bf v}^0|^2 dV = { m(m+1) \over 2m+3} \rho R_c
\ee
Putting all these ingredients together 
we find that (using definitions from the previous section) 
\be
s = { 2 (2m+3) Q_{m+1} \over (m+1)^2} \sqrt{ \Omega \over R_c^2} {\cal J}
\ee
which is in  agreement with the result, Eq.~(\ref{firsts}), of the previous
section. 

This final approach to the problem is obviously very elegant, 
and much less involved that the previous two calculations we have presented. 
Unfortunately, the surface integral approach relies upon the assumption of
incompressible flow. This assumption will not hold for stratified stars \citep{abney}, see Section~\ref{sec:compress}, 
or indeed stars with an internal magnetic field. Hence, we do not expect this method 
to be of much use for more complex (realistic) problem settings.


\section{Comparing different results}
\label{sec:compare}

Before we move on to consider the more complex problem of a compressible star, 
it is meaningful to compare and contrast the different results for the Ekman layer damping of r-modes
in the current literature. Let us do this in chronological order. That the 
viscous boundary layer would provide an efficient damping mechanism 
was first suggested by \citet{BU}. They estimated the associated 
damping rate by using the familiar plane-parallel oscillating plate result 
from \citet{landau}. Their result is not
too far from the  more 
detailed results derived in this paper, even though it does not account for the geometrical 
factors due to the sphericity of the star and the nature of the 
r-mode fluid flow. This is not very surprising since the 
geometric factors are of order unity as long as the thickness of the boundary layer is 
considerably smaller than the size of the star. At a local level, the boundary layer flow is
quite similar to that in the plane-parallel case.

In his comment on these first estimates, \citet{rieutord} extracted 
(\ref{firsts}) from Greenspan's work. As we have seen, this 
result has the relevant geometric factors etcetera, but it is based on 
a uniform density model.  \citet{owen} extended the standard method discussed in 
Section~\ref{sec:standard} to allow for non-uniform density models, and derived the boundary
layer-modified r-mode flow. Then, using the viscous shear tensor they arrive at the 
final result for the r-mode damping timescale,
\be
t_\mathrm{E} = - { 2 E \over \dot{E}} = { 2^{m+3/2} (m+1)! \over 2m\Omega (2m+1)!! {\cal I}_m } 
\sqrt{ 2 \Omega R_c^2 \rho_c  \over\eta_c  }  \int_0^{R_c} 
{ \rho \over \rho_c} \left( { r \over R_c} \right)^{2m+2} { dr \over R_c}  
\label{linds}\ee
where $\eta_c$ and $\rho_c$ are the viscosity coefficient and the density at the 
core-crust interface, respectively, and ${\cal I}_m$ is an angular integral analogous to our ${\cal J}$.
The integral on the right-hand side is essentially the kinetic energy of the r-mode.  
In order to compare this expression to the result quoted by Rieutord, we should 
take $\rho = \rho_c=\mathrm{constant}$ and work out the radial integral. This readily 
yields Eq.~(11) in \citet{rieutord}. Hence, the two results are consistent, as they should be.
To compare with our result, Eq.~(\ref{firsts}), we need to use the relation between 
the two angular integrals. One can show that
\be
\mathrm{Re}\ {\cal J} = { m (2m+3)!! \over 2^{m+2}}Q_{m+1} {\cal I}_m
\ee
Given this relation our result for the damping timescale
\be
t_\mathrm{E} = { (m+1)! \over 2 (2m+3) Q_{m+1} } 
\sqrt{ {\rho_c R_c^2 \over \eta_c \Omega}} { 1 \over \mathrm{Re}\ {\cal J}}
\ee
is also identical to Rieutord's result. 
From this comparison it is clear that any difference between the various 
detailed studies of the Ekman layer problem in the r-mode instability
literature is due to the chosen stellar model. 

To quantify the dependence of the damping rate on (say) the mass distribution in the star 
is obviously an important task.  \citet{owen} did this
by performing calculations for a number of realistic supranuclear equations of state. 
Here we will choose a more pragmatic approach, which we think provides
some useful insights. Let us first return to (\ref{linds}),     
but instead of taking $\rho=\rho_c$ we will, somewhat inconsistently, 
allow the two (constant) densities to be different. Then we see that, in the case of the $m=2$ mode,
\be
I = \int_0^{R_c}  {\rho \over \rho_c} \left( { r \over R_c} \right)^{6} { dr \over R_c}
= { \rho \over 7 \rho_c} \approx 3.41\times10^{-2} { M \over R^3 \rho_c}  
\ee
Now compare this result to a star described by an $n=1$ polytrope
for which
\be
\rho(r) = { M \over 4 R^2 r} \sin \left( { r \pi \over R} \right)
\label{poly}\ee
If we take $\rho_c=1.5\times10^{14}\ \mbox{g/cm}^3$ and  $R_c/R\approx 0.85$, which is a fairly typical value 
according to Table~1 in \citet{owen}, we find that
\be
I \approx 3.49 \times 10^{-2} { M \over R^3 \rho_c}
\ee
From this comparison we see that one can quite easily
bridge the difference between the constant density calculation and the 
results of \citet{owen}. As a further 
exercise, one can work out the integral by combining the 
data in Table~1 of \citet{owen} for $M$, $R$ and $R_c$
with the $n=1$ polytropic density profile. This leads to underestimates of 
the damping timescale results obtained by Lindblom {\it et al} for various equations of state by 
10-30\%.  This is due to the fact that realistic equations of state are 
slightly softer than the $n=1$ polytrope. Nevertheless, it shows that the polytropic model 
captures the main features of the problem. One should  keep in mind that the 
use of realistic equations of state for Newtonian neutron star models 
is dubious. For a given central density (say) the mass and radius 
often differ considerably from the corresponding results in 
general relativity. This means that it may not be entirely consistent to use
realistic equations of state in the present discussion. On the other hand, 
it is important to get an understanding of how the results change with 
the stellar parameters, and over what range the damping timescale
for an unstable r-mode may vary. We think that the above discussion led to some useful
 comparisons that provide useful steps 
towards this understanding. Possibly the most important conclusion is that
all current calculations are consistent.


\section{Including stratification and compressibility}
\label{sec:compress}

So far, we have discussed the Ekman layer problem for incompressible fluids.
Although this is not an unreasonable first approximation for neutron stars, it is well
documented that both compressibility and stratification due to either temperature or composition gradients 
may be important. Considering also the results of \citet{abney} for the spin-up problem, 
which suggest that the Ekman layer timescale is significantly altered by both 
stratification and compressibility, it is clear that we need to relax our 
assumptions. 

\subsection{Formulating the compressible problem}

We want to avoid the  assumption of uniform density and include
stratification and compressibility (while maintaining the calculation at ${\cal O}(\Omega)$ which means 
that the star is spherical). In a perturbed compressible, stratified star the equation of state is 
usually taken to have the form
\be
\Delta p = { p \Gamma \over \rho} \Delta \rho
\ee 
where $\Delta$ represents a Lagrangian variation and  $\Gamma$ is the adiabatic index. 
Written in terms of Eulerian perturbations ($\delta$) this means that, for our 
inviscid leading order flow we have
\be
\frac{\delta \rho^0}{\rho} = \frac{\dpe^0}{\Gamma p} -\mathcal{A}_s \xi_r^0
\ee  
where the Schwarzschild discriminant is
\be
\mathcal{A}_s = \frac{\rho^\prime}{\rho} -\frac{p^\prime}{\Gamma p}
\ee
(the primes represent radial derivatives). The Schwarzschild discriminant encodes information
regarding composition gradients etcetera \citep{reigol}

The question now is, how does this more complex equation of state 
affect the boundary layer calculation? At first sight one might think that 
the answer will be quite messy. Fortunately, this turns out not to be the case. 
Let us first consider the inviscid flow discussed in Section~\ref{sec:Ekman}.
Then we see that the introduction of the enthalpy was, indeed, a shrewd move. 
With the equations written in this way we do not need to worry about the 
relation between the perturbed density and pressure unless these variables are 
explicitly required. In our case, they are not. Thus we swiftly prove that 
the inviscid r-mode flow remains unchanged for a compressible model. 

The leading order boundary layer flow also requires little thought. 
As before we find $\tilde{v}^0_r = \delta \tilde{p}^0 =0  $. Given this, the remaining equations at this
level are identical to those from 
the uniform density calculation. Hence $\tilde{v}^0_{\theta} $ and $\tilde{v}^0_{\varphi}$  are obtained as in Section~\ref{sec:spherical}. 
Finally, from the equation of state we find,
\be
\frac{\delta\tilde{\rho}^0}{\rho} = \frac{\delta\tilde{p}^0}{\Gamma p} = 0   
\ee 

The complications associated with stratification and compressibility enter the problem at the 
level of the induced core flow. Expressed in terms of the enthalpy the  Euler equations (\ref{int1})-(\ref{induced}) 
remain unaltered, but the 
continuity equation is now given by
\be
i\sigma_0\,\frac{\dpe^1}{\Gamma p} - \frac{\rho g}{\Gamma p} v^1_r + 
\nabla \cdot {\bf v}^1 + \frac{i s \mathcal{A}_s}{\sigma_0} v^0_r = 0   
\label{conti}\ee
where we have prefered to use the perturbed pressure in order to emphasize the presence of the stratification. 
Inspection of the Euler 
equations, together with the fact that  $s \sim {\cal O}(\Omega^{1/2}) $, implies that 
$v^1_{\theta,\varphi} \sim {\cal O}(\Omega^{-1/2})$ and $\dpe^1 \sim {\cal O}(\Omega^{1/2})$. 
Hence, in the continuity equation  the second and third term are ${\cal O}(\Omega^{-1/2}) $ while
the first term is of order ${\cal O}(\Omega^{3/2}) $. At the same time, 
the fact that we are considering an r-mode flow means that the last term is also ${\cal O}(\Omega^{3/2}) $.
As we are working in the slow-rotation limit
 we keep only the former terms, and we get
\be
\nabla \cdot {\bf v}^1 -\frac{\rho g}{\Gamma p} v^1_r  = 0 
\label{conti2}
\ee
where $g=p^\prime/\rho$.
It is worth remarking that \citet{abney} make similar assumptions in their analysis of the spin-up problem.
At this point we can infer that 
stratification has no effect (at leading order) on the Ekman damping rate of a r-mode. This is in contrast 
to the situation for a general inertial mode (and the spin-up problem) for which the radial velocity component is 
of the same order as the angular ones. Then the last term in  (\ref{conti}) must be retained and 
stratification could be important. For our problem this turns out not to be the case, which may be a bit of a surprise.  

Our next task is to calculate the induced flow ${\bf v}^1 $. Asssuming the usual decomposition of
the velocity field we retain (\ref{indV}), i.e. 
\be
V_{m+1}^1 = \frac{is(m+1)}{2\Omega Q_{m+1} (m+2)} U^0_m -\frac{1}{m+2} W^1_{m+1}
\ee
At the same time the continuity equation gives,
\be
r\partial_r {W^1_{m+1} } + \left [ 1 - \frac{r}{R} \kappa \right ] W^1_{m+1} = 
(m+1)(m+2) V^1_{m+1}
\ee
where we have defined the dimensionless ``compressibility'',
\be
\kappa \equiv \frac{\rho g R}{\Gamma p}
\ee
Combining these equations we arrive at 
\be
r\partial_r {W^1_{m+1} } + \left [ m +2 - \frac{r}{R} \kappa \right ] W^1_{m+1} = 
\frac{is (m+1)^2}{2\Omega Q_{m+1}} U_m^0
\label{wnew}\ee
Once we solve this equation, we can replace (\ref{wrad}) in the analysis of the incompressible case
and infer the damping rate of the r-mode. The problem is that, whereas the $\kappa=0$ case is easily
dealt with, the radial variation of $\kappa$ in a realistic model will typically require a
numerical solution.  

\subsection{Case study: The $n=1$ polytrope}

To get an idea of how important the compressibility corrections may be we will consider 
the simple model problem of an $n=1$ polytrope. 
Before doing this, let us make a few comments on the work of \citet{abney}. In their analysis of the
spin-up problem they arrive at 
an equation corresponding to (\ref{wnew}).  They then make the assumption that $\kappa$ can be taken
as constant. Since $\kappa \gg 1$ in the region of the boundary layer the calculation simplifies considerably,
and one easily shows that the compressibility leads to an exponential suppression of the Ekman layer damping.
Working out the relevant algebra, we find a revised Ekman damping rate $\tilde{s}$ given by
\be
\frac{\tilde{s}}{s} \approx \frac{ \kappa^{2m+3}}{(2m+3)!}\left (\frac{R_c}{R} \right )^{2m+3} e^{- \kappa R_c/R}
\ee
For canonical values $m=2, R_c = 0.9 R $ and  $\kappa= 10$ (see below) we would have
\be
\frac{\tilde{s}}{s} \approx 0.1
\ee
This suggests that the Ekman layer damping 
is strongly suppressed in a compressible star. In fact, if we take $\kappa \approx 100$ as in \citet{abney}
we would have a truly astonishing result where the boundary layer dissipation was completely
irrelevant. The analysis is, however, wrong.

To see what the problem is, we can work out $\kappa$ for the polytrope (\ref{poly}). This leads to
\be
\kappa = { 1 \over x} \left( 1 - \pi \cot \pi x \right) \approx {1 \over 1-x } 
\ee
where $x=r/R$ and the last step follows from an expansion near the surface. As it turns out, this
is a good approximation throughout the star.  From this we see that while $\kappa r/R$ is indeed large
in the outer regions of the star (the divergence at the surface is due to the sound speed vanishing there), 
it vanishes at the centre and is certainly not large compared to $m+2$ in (\ref{wnew}) in the inner regions of the star. 
This is relevant for our analysis since we are trying to work out the induced flow in the core, not just the 
corrections in the thin boundary layer at the base of the crust.

Given the approximate expression for $\kappa$ it is straightforward to write down an
analytic solution to (\ref{wnew}). 
If we focus on the $l=m=2$ r-mode and express the right-hand side as $B x^3$, then the
solution is simply
\be
W^1 = {Bx^3 \over 7} \left( { 1 - 7x/8 \over 1-x} \right)
\ee  
This should be compared to the incompressible result $W^1_{\rm inc} = Bx^3 /7$.
Repeating the analysis following (\ref{wrad}) we find that
\be
\frac{\tilde{s}}{s} \approx { 1 - R_c/R \over 1- 7R_c/8R} 
\ee
This shows that the compressibility lowers the Ekman damping rate by a factor of two or so (for $R_c/R=0.9$).
Although this is far from irrelevant, it is certainly not as dramatic a suppression as the
results of \citet{abney} suggest. 

The above analysis is, as far as we know, the first attempt to account for the effect of
compressibility on the Ekman layer damping of r-modes. We have shown how the problem can be solved (up to quadrature), 
and then estimated the magnitude of the effect for a polytropic model. Would it now be useful to
consider 
an even more realistic model based on some tabulated equation of state? We think that the answer is no. 
There are two reasons for this. We have already discussed the well-known fact that one should be careful when using realistic
(relativistic) equations of state in a Newtonian calculation, and that it is 
difficult to make sure that one is comparing like with like. A second reason 
why we feel that a more ``detailed'' calculation may not be relevant is that we are still not considering the 
true physics problem. In a mature neutron star we should incorporate both superfluid components at
the crust-core interface and the magnetic field. We have shown that the Ekman layer damping in an 
incompressible normal fluid core differs from a compressible model by  a factor of a few. 
This is typical of the uncertainties associated with (say) the equation of state, and is perhaps the level 
of accuracy that we should expect to be able to achieve.


\subsection{Generalising the surface integral approach}
\label{sec:surf2}

To complete our discussion of the compressible problem it is relevant to 
consider how the surface integral approach discussed in Section~\ref{sec:surf} would have to 
be altered. 
Writing the Euler equations in terms of enthalpy, as in Section~\ref{sec:Ekman}, 
instead of pressure we have
for the inviscid flow:
\be
i\sigma_0 {\bf v}^0 + 2{\bf \Omega} \times {\bf v}^0 = - {\bf \nabla} \delta h^0
\ee
Similarly, the ${\cal O}(\nu^{1/2}) $ induced flow is described by
\be
i\sigma_0 {\bf v}^1  + 2{\bf \Omega} \times {\bf v}^1 = - {\bf \nabla} \delta h^1
+ s {\bf v^0}
\ee
We have seen that $[s, \delta h^1] \sim {\cal O}(\Omega^{1/2}) $ and 
$ {\bf v}^1 \sim {\cal O}(\Omega^{-1/2}) $.

Repeating the steps described in Section~\ref{sec:surf}  we obtain,
\be
s | {\bf v}^0 |^2 = \nabla \cdot [ {\bf v}^1 \delta \bar{h}^0 +  {\bf \bar{v}}^0 \delta h^1 ] 
- (\nabla \cdot {\bf v}^1 ) \delta \bar{h}^0 -(\nabla \cdot \bar{{\bf v}}^0) \delta h^1
\ee
The first three terms are ${\cal O}(\Omega^{1/2}) $. The last term is much smaller,
${\cal O}(\Omega^{5/2}) $, and can therefore be neglected. Transforming the first integral to
a surface integral in the standard way, we get at leading order,
\be
s \int  | {\bf v}^0 |^2 dV = \oint  ({\bf \hat{r}} \cdot {\bf v}^1 ) \delta \bar{h}^0 dS
- \int  ( \nabla \cdot {\bf v}^1) \delta \bar{h}^0 dV 
\label{sint2}
\ee
This is the generalisation of eqn.~(\ref{sint}) for the case of a stratified and compressible
fluid. The main difference from the incompressible result is the presence of the second term in the right-hand side. 
Since we know that, cf Eq.~(\ref{conti2}), 
\be
\nabla \cdot {\bf v}^1 = \frac{\kappa}{R} v^1_r
\ee
we see that it encodes the dependence on the compressibility. Since it is a volume integral, it is also clear that
the answer depends on the compressibility throughout the star not just in the boundary layer. 

\section{Summary}


We have revisited the problem of the damping of neutron star r-modes due to the
presence of an Ekman layer at the core-crust interface. This problem continues to be of 
interest since 
there is a  possibility that the gravitational-wave driven instability of the  
r-modes is active in rapidly spinning neutron stars, eg. in 
low-mass X-ray binaries. The Ekman layer is known to provide
one of the key damping mechanism 
for these oscillations. 

We reviewed various approaches to the problem and carried out an analytic 
calculation of the effects due to the Ekman layer for a rigid crust. 
Our obtained estimates support previous numerical results, 
and provide some useful insights into the problem. Of particular interest
should be the discussion of the breakdown of the ``standard'' solution to the problem 
associated with singularities in the velocity field in certain angular directions. 
Our comparison of previous results, which demonstrates that they are all consistent, 
is also useful. 

We extended the analysis to include
the effect of compressibility and 
composition stratification. This led to a demonstration that
 stratification is unimportant for the r-mode problem, but compressibility suppresses 
the damping rate by about a factor of two (depending on the detailed equation of state).
The physical interpretation is that the
compressibility counteracts the induced flow, and leads to a slightly less efficient
energy transport away from the boundary layer.
Our result corrects a similar analysis of \citet{abney} for the spin-up 
problem (associated with pulsar glitches). By assuming that the compressibility can be taken 
to be constant they predicted an exponential 
suppression of the Ekman layer damping. The reason why this assumption is not valid should be 
clear from our discussion.

To conclude, we think that this study clears the stage for work on more physically realistic models. 
The methods required to model Ekman layers in rotating neutron stars are now well understood. 
In particular, we understand the strengths and weaknesses of the different routes to the answer. 
In order to make our models more realistic we need to account for both the presence 
of magnetic fields and superfluid components. The effect of magnetic boundary layers have
been discussed by 
\citet{mend1} 
and \citet{kinney}. Their work shows that the magnetic field is of key importance.
There are also  many issues associated with superfluid neutron stars, ranging from the presence
of additional dynamical degrees of freedom to novel dissipation mechanisms. It is clear that, while we have made 
progress on the simplest Ekman layer problems, many challenges remain. 


\section*{Acknowledgements}

This work was supported by PPARC through grant number PPA/G/S/2002/00038.
NA also acknowledges support from PPARC via Senior Research Fellowship no
PP/C505791/1.

\section*{Appendix}

In this Appendix we provide the steps needed to
simplify the angular integral in Section~\ref{sec:standard}.
We need to work out
\be
{\cal J} = 2\pi \int_0^\pi \sin \theta  \bar{Y}_{m+1}^m \left\{ 
{ i m(m+1) \over 2} \left[ { 1 \over \bar{\lambda}_1} - 
{ 1 \over \bar{\lambda}_2}  \right] Y_m^m
- AR_c { \partial_\theta \bar{\lambda}_1 \over \bar{\lambda}_1^2}
- BR_c { \partial_\theta \bar{\lambda}_2 \over \bar{\lambda}_2^2}\right\} 
d\theta 
\ee
To do this we use
\be
\int_0^\pi \sin \theta  { 1 \over \bar{\lambda}_1}  Y_m^m\bar{Y}_{m+1}^m
d\theta = - 
\int_0^\pi \sin \theta  { 1 \over \bar{\lambda}_2}  Y_m^m\bar{Y}_{m+1}^m  
d\theta
\ee
and
\be 
\int_0^\pi \sin \theta 
A { \partial_\theta \bar{\lambda}_1 \over \bar{\lambda}_1^2}
  \bar{Y}_{m+1}^m d \theta = 
\int_0^\pi \sin \theta 
B { \partial_\theta \bar{\lambda}_2 \over \bar{\lambda}_2^2}
  \bar{Y}_{m+1}^m d \theta
\ee 

In other words, we have
\be 
{\cal J} = 4\pi \int_0^\pi \sin \theta  \bar{Y}_{m+1}^m \left\{ 
{ i m(m+1) \over 2 \bar{\lambda}_1}  Y_m^m
- AR_c { \partial_\theta \bar{\lambda}_1 \over \bar{\lambda}_1^2}
\right\} 
d\theta
\label{jcal} 
\ee

We can simplify this further in the following way;
\begin{eqnarray}
\int_0^\pi \sin \theta  \bar{Y}_{m+1}^m 
AR_c { \partial_\theta \bar{\lambda}_1 \over \bar{\lambda}_1^2}
d\theta &=& - \int_0^\pi \sin \theta  \bar{Y}_{m+1}^m 
AR_c  \partial_\theta \left( {1 \over  \bar{\lambda}_1}\right)
d\theta =
\int_0^\pi  {1 \over  \bar{\lambda}_1 }  \partial_\theta \left( \sin \theta 
\bar{Y}_{m+1}^m AR_c \right) d\theta = \nonumber \\
&=& { i \over 2} \int_0^\pi { \sin \theta  \over \bar{\lambda}_1}
\left\{ m(m+1) \bar{Y}_{m+1}^m Y_m^m - { m^2 \over \sin^2 \theta} 
\bar{Y}_{m+1}^m Y_m^m -(\partial_\theta  \bar{Y}_{m+1}^m)(\partial_\theta 
Y_m^m) \right. 
\nonumber \\ 
\nonumber \\
&+& \left. { m \over \sin \theta} \partial_\theta (  
\bar{Y}_{m+1}^m Y_m^m )  
\right\} d\theta
\end{eqnarray}
This way we arrive at Eq.~(\ref{Jint}) in Section~\ref{sec:standard}.



\end{document}